\documentstyle[12pt]{article}

\newcommand{\half}{\frac 1 2 }

\newcommand{\ie}{{\em i.e.} }

\newcommand{\noi}{\noindent}

\newcommand{\pref}[1]{(\ref{#1})}
\newcommand{\kah}{K\"ahler }

\newcommand{\ee}{\end{eqnarray}}

\newcommand {\be}[1]{
      \begin{eqnarray} \mbox{$\label{#1}$}  }

\gdef\journal#1, #2, #3, 1#4#5#6{{#1~}{#2} (1#4#5#6) #3} 
\gdef\ibid#1, #2,1#3#4#5{{\bf #1} (1#3#4#5) #2}

\begin{document}
	\vskip 5mm
	\begin{center}
{\Large\bf Classical phase space and statistical \\ 
mechanics of identical particles }
	\end{center}
	
\vspace* {-41 mm}
\begin{flushright}
  USITP-00-02 \\
  OSLO-TP 2-00 \\
  March-2000 \\
\end{flushright}

\vskip 25 mm
\centerline{\bf  T.H. Hansson$^{1}$, S.B. Isakov$^{2}$, J.M.
Leinaas$^{3}$,  and  U. Lindstr{\"o}m$^{4}$ }

\medskip
{\centerline{\it ${}^{1,4}$ Department of Physics, University of
Stockholm, Box 6730,\\ S-11385 Stockholm, Sweden }}
{\centerline{\it ${}^{2,3}$Department of Physics,  University of Oslo,
P.O. Box 1048 Blindern, N-0316 Oslo, Norway}}

\date{October 1998}

\vskip 5 mm
\vskip 20mm

\centerline{\bf ABSTRACT}
\vskip 3mm
Starting from the quantum theory of identical particles, we show how 
to define a classical mechanics that retains information about the 
quantum statistics. We consider two examples of relevance for the 
quantum Hall effect: identical particles in the lowest Landau level, and 
vortices in the Chern-Simons Ginzburg-Landau model. In both cases the resulting {\em 
classical} statistical mechanics is shown to be a nontrivial 
classical limit of Haldane's exclusion statistics. 
\vfill
\noindent
e-mail: $^1$ hansson@physto.se, $^2$
serguei.isakov@fys.uio.no, 
$^3$ j.m.leinaas@fys.uio.no, $^4$ ul@physto.se 
 \\

\noindent
$^2$ Present address: NumeriX Corporation, 546 Fifth Avenue, New York, NY
10036, USA \\

\noindent $^{1,4}$ Supported by the Swedish Natural Science Research
Council
\\
\noindent
$^3$ Supported by the Norwegian Research Council \\

\newpage
\noindent{\large\bf 1. Introduction }
\renewcommand{\theequation}{1.\arabic{equation}}
\setcounter{equation}{0}
\vskip 5mm
Particle statistics is usually considered to be a quantum effect.
It is expressed through the symmetry of the wave function of a system of
identical particles and does not appear, normally, as an interaction in
the Hamiltonian. In two space dimensions it can be represented as a
special kind of interaction, but being of the Aharonov-Bohm type, it
does not give rise to any force on the particles.

Thus, at the level of classical trajectories of individual particles,
there is no  difference between identical and non-identical
particles. There is however one place in the classical description of
particles where their indistinguishability  is important,
namely in the  statistical mechanics. There the
trajectories of individual particles no longer matters, but the volume of
the available phase space is important for thermodynamical quantities.
Indistinguishability is introduced by dividing the phase space volume
of $N$ non-identical particles with the factor $N!$. This reduction is
essential to give the correct expression for the entropy and thus to 
resolve Gibbs' paradox.

The reduction in phase space is readily understood. If the particles are
indistinguishable all configurations that can be related by a permutation
of the particles correspond to one and the same {\em physical}
configuration. This single configuration for identical particles is then
represented as
$N!$ different configurations in the case of distinguishable particles.
The configuration space of indistinguishable point particles is
therefore derived from the space of distinguishable particles by an
identification of equivalent points. 

The identification of points implies that the configuration space of a
system of identical particles is not everywhere a smooth manifold, there
are singularities corresponding to the situation where two or more
particles occupy the same point in space. Such a point is a geometrical
singularity, a point of infinite curvature.  For the phase space the situation
is similar, the identification of points introduce singularities, although in
general these singularities do not have the same simple geometrical
interpretation as in configuration space.

The quantum description of identical particles can be introduced in
terms of wave functions, or alternatively in terms of path integrals,
defined on the configuration space with identifications
\cite{Laidlaw71,Leinaas77}. The presence of singularities then are important,
since it divides the continuous paths into different classes, depending on how
they evolve around the singularities. Such classes can be associated with
different phase factors. There is only one characteristic phase factor for each
system of identical particles, corresponding to an exchange of two 
particles, and this factor identifies the
statistics. Viewed in this way, the statistics parameter
associated with the particles labels inequivalent quantizations of
the classical system. Thus, the statistics parameter appears in the
quantization of the system and is not present in the classical 
description of the particles.

In this paper we will to discuss an alternative approach to the classical
description of identical particles. This does not mean that we consider the
standard description of point particles referred to above as being in any sense
incorrect. However, we would like to stress that starting from the
quantum theory there are different possibilities for describing the
corresponding classical system, and we would like to examine one where the
statistics parameter is present also at the classical level. As discussed in
the paper we may view this as a non-standard way of taking the classical limit.

The way we introduce the classical description is to consider, in a general
form, a coherent state representation of the quantum system. We assume
the coherent states to be determined by a set of particle coordinates, and we
further assume the time evolution (in the low
energy regime) to a good approximation  to be described simply by the  
motion of these coordinates. There is a manifold defined by the set of
possible coordinates and a natural phase space structure inherited from the
full quantum description. This phase space is a smooth manifold, even when the
particle coordinates coincide, and the reduction corresponding to the factor
$1/N!$ does not have to be introduced by hand, but appears naturally when
calculating the phase space volume.

To clarify this idea, we consider the case of a harmonic oscillator coherent
state representation of a system of identical point particles in some detail. We
show how the classical description introduced in this way
distinguishes between bosons, fermions, and in general 
anyons \cite{Leinaas77,anyons}, and we calculate the
available phase space volume for the case of $N$ identical particles in a
finite volume. The classical statistics parameter is then identified as
the phase space volume occupied by each of the particle present in the
system. Viewed in this way the description has the character of a
classical analogue of the quantum exclusion statistics introduced some
time ago by Haldane \cite{Haldane91}. We examine this correspondence in some
detail by considering the  statistical mechanics  of our classical system.

The description we use is not restricted to systems of point
particles. We illustrate this by considering vortex solutions of
the Chern-Simons Ginzburg-Landau theory. The manifold defined by
the $N$-vortex configurations has a natural phase space structure,
and although this  can not be fully determined, the phase
space volume can be calculated and the statistics parameter 
identified. This particle description of vortices is closely
related to a description of vortices in the (relativistic) abelian
Higgs model previously discussed by Samols, Manton and others
\cite{Samols92,Manton93} although in their case the vortex manifold is identified
as a  configuration space rather than as a phase space.       

\vskip 8mm
\noindent{\large\bf 2. Classical phase space from the quantum description}
\renewcommand{\theequation}{2.\arabic{equation}}
\setcounter{equation}{0}
\vskip2mm

\def\zb{\overline z}
\def\zpb{\overline z'}
\def\ad{a^{\dagger}}
\def\bz{{\bf z}}
\def\bzb{{\bf \zb}}
\def\bm{{\bf m}}

\newcommand{\bra}[1]{\langle #1 |}
\newcommand{\ket}[1]{|#1\rangle}
\newcommand{\bracket}[2]{\langle #1|#2 \rangle}
\newcommand{\dop}[1]{D(#1)}

\newcommand{\za}[1]{z_{#1}}
\newcommand{\zab}[1]{\overline z_{#1}}

\newcommand{\pza}[1]{\partial_{z_{#1}} }
\newcommand{\pzab}[1]{\partial_{\overline z_{#1}} }

\newcommand{\zz}[2]{\overline z_{#1}z_{#2}  }
\newcommand{\braket}[2]{\langle #1 | #2 \rangle}
\newcommand{\nN}{{\cal N}}

In this section we consider a general quantum system and a subset of
states $\ket {\psi_x}$, which is indexed by a set of coordinates
$x=\{x_1,x_2,...,x_N\}$. These may be the coordinates of a system of
(identical) particles or the coordinates of an $N$ soliton configuration,
but we do not have to be more specific at this point. 
We only assume that the wave
function evolves smoothly with a change of these coordinates, and that it
is symmetric under an interchange of any pair of the $N$ coordinates. We
furthermore assume, that in the regime of interest (typically  at
low energies),  the time evolution of the system, to a good approximation, 
can be described (up to a phase factor) as a time evolution of 
the coordinates only. 
This means that it makes sense to
consider the restricted (constrained) system where the evolution of the
system is projected to the manifold defined by the normalized states $\ket
{\psi_x}$.  Since the physical states correspond to rays in the Hilbert
space,
\ie to state vectors defined up to a complex factor, we consider the
classical $N$-particle space $\cal M$, derived 
from the quantum description, to be defined by the normalized states
$\ket {\psi_x}$ only up to such a phase factor.
It is the phase space structure of the space
$\cal M$ which will be of importance for our discussion. 

The Schr{\"o}dinger equation of the quantum system can be
derived from the Lagrangian, 
\be{glag1}
L=i\hbar \bracket \psi {\dot \psi} - \bra\psi  H \ket\psi \; ,
\ee
where $H$ is the Hamiltonian of the system, and the Lagrangian of the
constrained system is obtained from this  by restricting $\ket\psi$
to the subset of states $\ket{\psi_x}$. Expressed in terms of the coordinates
$x$, it has the generic form
\be{glag2}
L= \dot x_iA_i(x)-V(x) \; ,
\ee
where $A_i$ is the Berry connection \cite{Berry84} 
\be{gpot}
A_i=i\hbar \braket {\psi_x} {\partial_i{\psi_x}}\; ,
\ee
$\partial_i$ denotes the partial derivative with respect to 
$x_i$,\footnote{
We use a shorthand notation by treating $x_i$ as a
single parameter. In reality the phase space space for each particle will
be multi-dimensional.} 
and the potential
$V$ is  the expectation value of the Hamiltonian in the state
$\ket {\psi_x}$. The equation of motion derived from the Lagrangian is,
\be{geom}
f_{ij}\dot x_j=\partial_{i} V \; ,
\ee
with
\be{gfs}
f_{ij}=\partial_iA_j-\partial_jA_i \; .
\ee
Under the general condition that $f_{ij}$  is an everywhere invertible
matrix (which in particular means that the space $\cal M$ is
even-dimensional), a Poisson bracket can be defined and a symplectic
structure introduced on $\cal M$. The bracket has the form \cite{Jackiw1}
\be{brack}
\{A,B\}=(f^{-1})_{ij}\,\partial_iA\,\partial_jB \; ,
\ee
and the equation of motion  can then be written as
\be{geom2}
\dot x_i=\{x_i,V\} \; .
\ee
The corresponding symplectic form is,
\be{spf}
\omega=-\half f_{ij}\,dx_i\wedge dx_j \; ,
\ee
and in particular this  determines the phase space volume. Thus,
under the general conditions mentioned, a classical phase space can be
derived from the quantum description. Note, however, that it is a
generalized phase space in the sense that no configuration space has been
identified. 

The symplectic structure of the manifold $\cal M$ has a simple geometric
interpretation. It is defined as the imaginary part of the scalar product in the
tangent space of $\cal M$, which is obtained by projection from the
Hilbert space of the full quantum system and can be written as
\be{gfs2}
f_{ij}=-2\hbar \Im \{\braket{D_i\psi_x}{D_j\psi_x}\} \; ,
\ee
with $D_i$  the projected derivative,
\be{proj}
\ket{D_i\,\psi_x}=\ket{\partial_i\,\psi_x}-\ket{\psi_x}\braket
{\psi_x}{\partial_i\,\psi_x} \;.
\ee
Written in this form, it is manifest that the symplectic form, defining 
the classical kinetic energy, only depends on the properties of the
projected subspace. 
The real part of the scalar product gives another, related structure on
$\cal M$, which can be interpreted as a metric \cite{Provost80},
\be{gmet}
g_{ij}=2\hbar \Re\{ \braket{D_i\psi_x}{D_j\psi_x}\}\;.
\ee
This construction provides a natural way to introduce a metric on 
the phase space, and makes it possible to discuss its geometry.

At this point we will introduce an assumption about the geometrical
structure of $\cal M$, which leads to a simplification in the discussion
to follow. We assume $\cal M$ to be a K{\"a}hler manifold. This has
several technical implications, but we will only use that
there is a complex structure on $\cal M$, such that the symplectic and
metric structures referred to above are the antisymmetric and symmetric
part of the same complex K{\"a}hler metric. In terms of complex coordinates on
$\cal M$, we then get the following expressions for the
symplectic form and the metric,
\be{gkh}
\omega=-f_{\zb_iz_j}d\zb_i\wedge dz_j \; , \nonumber \\
ds^2=-2i f_{\zb_iz_j}d\zb_idz_j \; ,
\ee
with
\be{gfs3}
f_{\zb_iz_j}&=&\partial_{\zb_i}A_j-\partial_{z_j}A_{\bar i} \; .
\ee
This tensor can further be expressed as
\be{Kahler}
f_{\zb_iz_j}= i\partial_{\zb_i}\partial_{z_j}K(z,\zb)\; ,
\ee
where $K(z,\zb)$ is the K{\"a}hler potential. 

The condition that $\cal M$ is a K{\"a}hler manifold is satisfied when the
state vectors which define this manifold are, up to normalization, analytic
functions of
$z_i$,
\be{gnorm}
\ket{\psi_z}=\nN(\zb,z) \ket{\phi_z} \; ,
\ee
where $\ket{\phi_z}$ denotes the analytic part of the state vector and
$\nN(\zb,z)$ is the normalization factor. The vector potentials are then
given by
\be{gpot2}
A_i&=&-i\hbar\partial_{z_i}\ln{\overline\nN(\zb,z)}\; , \nonumber \\
A_{\bar i}&=&i\hbar\partial_{\zb_i}\ln\nN(\zb,z)\; ,
\ee 
and the K{\"a}hler potential is
related in a simple way to the normalization factor,
\be{gkh2}
K(\zb,z)=\hbar\ln |\nN (\zb,z)|^{-2} \; .
\ee 


\vskip 8mm
\noindent{\large\bf 3. Coherent states of identical particles }
\renewcommand{\theequation}{3.\arabic{equation}}
\setcounter{equation}{0}
\vskip2mm

We  now illustrate the general discussion
by considering coherent states of the one dimensional harmonic 
oscillator, or equivalently, charged particles moving in two 
dimensions in the presence of a strong magnetic field that restricts 
the available states to the lowest Landau level.
In this example we can explicitly derive the metric and 
symplectic structure, and show that they can be obtained 
from a K{\"a}hler potential. 

We first define the coherent states for bosons and fermions
and derive the corresponding classical mechanics. There is no 
unambiguous way to define coherent states for anyons, but we will 
use a construction which is very natural in this context, and again 
study the corresponding classical mechanics. 

Since we are  interested in the statistical mechanics, we also want to 
calculate the N-particle phase space volumes in the different cases. 
For this, it is necessary to start with a finite volume, and then take 
the thermodynamic limit. There are two obvious ways to confine the 
system, either by a potential, or by restricting the motion to a 
compact surface. In this section we shall consider the latter and 
study particles moving on a sphere. The case of a harmonic confining 
potential is treated in the Appendix.

\vskip 4mm
\noi{\bf A. Bosons and fermions in the plane}

We shall use the notation of ref.~\cite{pere1} and define a coherent 
state by translations of a minimum uncertainty reference state $\ket 
{ 0}$.
The  translation operators, $D(z)$, form a unitary
and irreducible representation of the 
Heisenberg-Weyl group, and in the following we shall  use the 
following explicit representation in terms of creation and 
annihilation operators,
\be{Ddef}
D(z) = e^{z\ad - \zb a} = e^{-\half\zb z}e^{z\ad}e^{-\zb a} \; ,
\ee
where $[a,\ad]=1$, and $z$ is a dimensionless complex coordinate.
In addition to the obvious relations $D(z)^{\dagger} = D(-z)$ and 
$D(0)=1$, we  shall need the following multiplication rule,
\be{dmult}
\dop{z_{1}}\dop{z_{2}} = e^{-\half(\zab 1\za 2 - \zab 2 \za 1) }\dop{\za 
1 + \za 2 } \; .
\ee
The coherent states  are now defined by, 
\be{csdef}
\ket z = D(z)\ket {0} =  e^{-\half\zb z}e^{z\ad}\ket{0} \; .
\ee
with a reference state $\ket{0}$ which is annihilated by $a$. 
For convenience we shall use a
notation where the normalized coherent  states are labeled by
$z$ only, although the normalization factor also  depend on $\zb$. This is to
distinguish the coherent states from the  position eigenstates $\ket {z,\zb}$,
and should lead to no confusion. From \pref{dmult} we immediately get  the
overlap between two coherent  states,
\be{overlap}
\braket {\za 1}{\za 2} = \bra 0 D^{\dagger}(\za 1)\dop{\za 2} \ket 
0 = \bra 0 \dop{-\za 1}\dop{\za 2 }\ket 0 = e^{-\half (\zz 1 1 + \zz 2 
2) + \zz 1 2} \; .
\ee
An unsymmetrized basis of N-particle coherent states is defined by
\be{ncs}
\ket{{\bf z}} = \ket{\za 1, \za2, \ldots \za N } &=&   D_{1}(\za 1) D_{2}(\za 2) 
\ldots D_{N}(\za N)\ket{0,0,\ldots 0} \nonumber\\
&&\\
&=& e^{\left({-\half\sum_{i=1}^{N} \zz i i }\right)} 
e^{\left({\sum_{i=1}^{N}
\za i 
\ad_{i} }\right)} \ket{\bf{0}} \; .
\nonumber \ee
Normalized N-particle coherent Bose and Fermi states are symmetric and 
antisymmetric linear combinations respectively,
\be{symcs}
\ket{\bz,\pm} = \ket{\za 1, \za2, \ldots \za N }_{\pm} &=& \nN (\za i,\zab i) \frac 1 
{\sqrt N!} \sum_{P}\eta_{P}^{\pm}  e^{ \za {i_{P}}
\ad_{i} } \ket{\bf{0}} \; ,
\ee
where sum is over permutations, $P$, and the sum in the exponent over the 
index i is suppressed. 
The permutation factor, $\eta_{P}$,  equals 1 for bosons and $\pm 1$ for 
fermions depending on whether the permutation is even or odd. Note 
that all dependence on $\zab i$ is in the normalization factor $\nN$, 
which is given by,
\be{norm}
|\nN (\bz,\bzb)|^{-2} = \sum_{P,P'}\eta_{P}\eta_{P'}\frac 1 {N!} 
  \bra{{\bf 0}} e^{\zab{j_{P'}}a_{j} }  e^{\za {i_{P}}
\ad_{i} } \ket{{\bf 0}} = \sum_{P}\eta_{P}e^{\zab{i_{P}}\za i } \; .
\ee

Following the general discussion in the previous section, we write the 
classical phase space Lagrangian (2.1) for the N-body system as, 
\be{nlag}
L(\bz , \bzb) = \bra{{\bf z },\pm}i\hbar\partial_{t} - \hat H \ket{{\bf 
z},\pm}\; ,
\ee
where $\hat H$ is the quantum Hamiltonian. In the following we shall 
use the harmonic oscillator as a simple illustration, \ie we take,
\be{ho}
\hat H = \hbar\omega\sum_{i=1}^{N}(\ad_{i}a_{i} + \half) \; .
\ee
Using \pref{nlag}, \pref{ho} and \pref{symcs} we get,
\be{nlag2}
L(\bz , \bzb) =  i\hbar(\dot{\zab i}\pzab i \ln \nN - \dot{\za i}\pza i 
\ln{\overline \nN} ) - \hbar\omega\za i \pza i \ln |\nN|^{-2}  - \half
N\hbar\omega\; .
\ee
By varying with respect to $\zab i$ one easily verifies that the equation of motion 
is  that of a harmonic oscillator, \ie $\dot\za i =- i\omega\za i$. We can rewrite 
\pref{nlag2} on the standard form (2.2),
\be{nlag3}
L(\za i, \zab i) = \half \left( A_{\zab i}
\dot{\zab i}  + A_{\za i}  \dot{\za i}\right) - V(\za i, \zab i) \; , 
\ee and using the phase convention $\nN=\overline \nN$ the potentials 
$A_{z}$ and $A_{\zb}$ and $V$ can all be obtained from the K\"ahler
potential \pref{gkh2},
\be{pot}
V(\bz , \bzb)  &=& \omega \za i \pza i K(\bz , \bzb)\; , \nonumber\\
A_i(\bz , \bzb)&=&\frac{i}{2}\partial_{z_i}{ K(\bz , \bzb)}\;
, \\ A_{\bar i}(\bz
,\bzb)&=&-\frac{i}{2}\partial_{\zb_i}K(\bz , \bzb)\; ,
\nonumber\ee
with
\be{kah}
K(\bz , \bzb)   =\hbar \ln |\nN(\bz , \bzb)|^{-2}   
= \hbar \sum_{P}\eta_{P}e^{ \zab{i_{P}} \za{i} } \; .
\ee
To get some insight into the meaning of these expressions we first 
analyze the two body case. Expressed in center of mass and relative 
coordinates, $Z=\half(\za 1 + \za 2)$ and $z=\za 1 - \za 2$, the \kah 
potential reads,
\be{twopot}
K(Z,z,\overline Z,\zb)_{\pm} = 2\overline Z Z + \ln\left[ e^{\half \zb z} \pm 
   e^{-\half \zb z}\right] \; ,
\ee
where $\pm$ refers to bosons and fermions respectively. The 
corresponding Lagrangians for the relative coordinate are now obtained 
using \pref{nlag3} - \pref{twopot}
\be{twolag}
L_{B}(\zb , z) &=& \frac {i\hbar} 4 (\zb\dot z - \dot{\zb} z)\tanh{\frac
{\zb z}  2} - 
   \frac {\hbar\omega} 2 \zb z \tanh{\frac {\zb z} 2} \; , \\
L_{F}(\zb , z) &=& \frac {i\hbar} 4 (\zb\dot z - \dot{\zb}z ) \coth{\frac
{\zb z}  2} 
  - \frac {\hbar\omega} 2 \zb z \coth{\frac {\zb z} 2} \; .\nonumber 
\ee
Note that fermionic Lagrangian is singular in the limit of small 
$r = \sqrt{ \zb z}$, 
\ie when the particles come close together. This is however of 
no physical significance, since the singular piece is a total time 
derivative,
\be{sing}
\lim_{r \rightarrow 0} L_{F} = \frac {i\hbar} 2 (\zb\dot z - 
\dot{\zb}z ) \frac 1 {\zb z} = \frac {i\hbar} 2 \partial_{t} \ln(z/\zb) 
 \; ,
\ee
which can be absorbed in $\nN$ as a pure phase (relaxing the reality condition),
or equivalently, as a  pure gauge term in $A_{z}$ and $A_{\zb}$.  From
\pref{Kahler} we get for the symplectic two form,
\be{bfmet}
f^{B}_{\zb z} &=& \frac{i\hbar} 2  \left( \tanh\frac {\zb z} 2 +
\frac {\zb z}  {2\cosh^{2}\frac {\zb z} 2  } \right) \stackrel
{r\rightarrow
\infty}\rightarrow \frac{i\hbar} 2 \; ,  \\  
f^{F}_{\zb z} &=& 
\frac{i\hbar} 2
\left(
\coth\frac {\zb z} 2 - \frac {\zb z}  {2\sinh^{2}\frac {\zb z} 2  }
\right) \stackrel {r\rightarrow \infty}\rightarrow   \frac{i\hbar} 2 \; ,
\nonumber
\ee
where $z=re^{i\phi}$.  Note that although the K{\"a}hler potential is singular in the 
fermion case, the metric is well defined.

In both cases, in the limit of large 
$r$,  we retain the naive flat metric, $ds^{2} = \hbar\; dz d\zb$. Since we refer
to relative coordinates this expression is reduced by a factor 2, as
compared to the case of a  single particle\footnote{ Our normalization is
such, that for a harmonic oscillator  hamiltonian with $m=\omega=1$, $z$ is
related to the  usual coordinates and momenta by, 
$z = (x + ip)/\sqrt{2\hbar}$, so $\hbar dzd\zb = [(dx)^{2} + 
(dp)^{2}]/2$.}.
(Note that the appearance of $\hbar$
in these classical expressions is due to the use of the
dimensionless coordinate $z$. As a more natural variable in the classical
description we may use the dimensional coordinate
$\tilde z=\sqrt{\hbar}z$ which then would remove $\hbar$ from the
expressions.) 

The more interesting 
limit is that of small r,
\be{metlim}
ds^{2} \stackrel {r\rightarrow  0} \rightarrow \hbar [\rho^{2} 
d\theta^{2} + d\rho^{2} ]& & \ \ \  (bosons) \; , 
\\
ds^{2} \stackrel {r\rightarrow  0} \rightarrow \frac \hbar 3 [\rho^{2} 
d\theta^{2} + d\rho^{2} ] & &\ \ \  (fermions) \; , \nonumber
\ee
where we have changed variables to $\rho=r^{2}/2$ and $\theta = 2\phi$. We
note that in these variables the metric has the standard flat-metric form in
polar coordinates. Thus, the new angular variable is restricted to an
interval of $2\pi$ (for $\rho=0$ to be a regular point), and therefore
$\phi$ is restricted to an interval of $\pi$. This has implications also for
large separation of the particles, where the space of relative coordinates has
the geometry of  a cone rather than that of a
plane. This is similar to the situation for the configuration space of two
identical particles when this is constructed by identification of physically
equivalent points \cite{Leinaas77}. However, in the present case the space is
geometrically smooth for small separation and the reduction in  volume
(essentially by a factor 2) compared to that of non-identical particles
appears naturally from the metric of the space and not through the
identification (by hand) of equivalent points.

The phase space of bosons and fermions has the same flat metric for large
separation. However, for small separation it is smooth but different in
the two cases. This leads to a correction to the phase space volume coming
from the short-distance behaviour which is different for bosons and
fermions. If we fix the maximal separation, $R$ \footnote{
R is again measured in dimensionsless units, just as z},
of the particles, the
volume of the interior of the selected region is determined from the form
of the potential
$A_z$ and $A_{\zb}$ on the boundary alone. With the description still
restricted to relative coordinates, the phase space volume becomes\footnote{
The
phase space volume given by
\pref{2partvolume} is identical to the Berry phase associated with the
interchange of the two particles. The close relation between phase space
volume and Berry phase is the analogue of the two well-known aspects of
quantum statistics: the symmetry of the wave function on one hand and the
Pauli exclusion principle on the other.}
\be{2partvolume}
V=-\int f_{\zb z}d\zb\wedge dz=-\left[\oint A_zdz+\oint A_{\zb}d\zb\right] .
\ee
This gives in the two cases 
\be{bose-fermi}
V_B=\half\hbar\pi R^2\; ,\;\;\;V_F=\half\hbar (\pi R^2-2\pi) .
\ee
The difference in phase space volume is a manifestation of the the difference in
statistics in the classical description. This is readily understood from a
semiclassical description where the number of states in a (single
particle) phase space is identical to the volume divided by $h$. (The factor
$1/2$ in \pref{bose-fermi} follows from the fact that we refer to relative
coordinates, with the angular integration in \pref{2partvolume} running from
$0$ to $\pi$. The volume of one particle-space with the position of the second
particle fixed would not include this factor.) In the following we will simply
take the reduction in phase space volume due to the presence of the other
particle as defining the statistics parameter in the classical description.
We will then examine how the phase space volume of an
$N$-particle system depends on this parameter, not only for bosons and fermions
but also for intermediate values of the statistics parameter.

\vskip 4mm
\noi{\bf B. Anyons in the plane}

It is well known that the harmonic oscillator coherent states
states are identical to the maximally localized states of charged
particles in a  magnetic field   projected to the lowest
Landau level (LLL). The translation from harmonic  oscillator to particle
in the LLL is as follows
\be{LLL}
a &=& \frac i {\sqrt{2}}( \tilde\Pi_{x} - i\tilde\Pi_{y})\ell \; ,  \\
z &=& \frac 1 {\sqrt{2}} (R_{x} - iR_{y}) \frac 1 \ell \; , \nonumber
\ee
where (in the symmetric gauge) $\tilde\Pi_{i} = p_{i} -\half
eB\epsilon_{ij}\hat x_{j}$ are  the generators of magnetic translations, 
$R_{i}$ the guiding  center coordinates, and $\ell = (\hbar/eB)^{\half}$
the magnetic length.  This relation  just expresses 
that the configuration space of charged particles in the LLL is 
mathematically equivalent to the phase space of a particle in one 
dimension. The re-interpretation of the coherent states as describing
particles in the lowest Landau level is helpful in two respects. We can in
a simple way generalize the coherent state representation of bosons and
fermions to fractional statistics, \ie to anyons in the lowest Landau
level. We can also more easily introduce a finite volume and take the
correct thermodynamic limit. Although there is no natural way to
restrict particle motion in one dimension to a finite phase space, charged
particles moving in a finite area  penetrated by a constant magnetic
field, makes perfect sense. In particular we can (in theory) study anyons
moving on a compactified space, like a sphere. This  problem will be
studied below, but first we generalize  our coherent state formalism to
the case of fractional statistics. 

In complex coordinates, an N-body anyon wave function has the form,
\be{anywf}
\Psi^{\nu} ({\bf z},{\bf\zb}) = \prod_{i<j}\left(\frac {\zab i - \zab j} {\za 
i -\za j} \right)^{\nu/2} \Psi_{B} ({\bf z},{\bf\zb}) \; ,
\ee
where $\Psi_{B}$ is a totally symmetric function. In general very 
little is known about anyonic energy eigenstates for
$N>2$. Exceptions are the LLL anyon states in a magnetic
field which are of the form \cite{anyonstates},
\be{anylll}
 \Psi^{\nu}_{\bf m}({\bf z},{\bf\zb})= \prod_{i<j}(\zab i - \zab j)^{\nu}  
 e^{-\half {\bf z\zb} } S_{{\bf m}}({\bf \zb}) \; ,
\ee
where ${\bf m} = (m_{1},\ldots m_{N})$, $m_{i}$ integer, and 
\be{sympol}
S_{{\bf m}}({\bf z}) =  \nN_{{\bf m}}{\cal S} \prod_{\bf m}\za 
 i^{m_{i}} \; .
\ee
${\cal S}$ is the symmetrization operator and $ \nN_{{\bf m}}$ a 
normalization constant. We now recall \cite{Kjonsberg97} that the fermion and 
boson coherent states in \pref{symcs}, up to a normalization factor, 
is nothing but the projection on the lowest Landau level of the 
appropriately (anti)symmetrized position eigenstates,\footnote{
This can easily be verified by explicit calculation, and it is also 
natural since the coherent states are the minimum uncertainty states 
centered around $\bz$. }
\be{projdef}
\ket{\bz ,\pm} =  C_{\pm}(\bz, \bzb) P_{LLL}\ket{\bz ,\bzb ,\pm} \; ,
\ee
which implies that any bosonic of fermionic N-body wave function in 
the lowest Landau level can be expressed as
\be{boslll}
\Psi^{\pm}({\bf z},{\bf\zb}) = \bra{\bz,\bzb , \pm}P_{LLL} \ket\Psi  =
C_{\pm}(\bz, \bzb)^{-1} \bracket{\bz,\pm} \Psi \; .
\ee
In particular, if we are given a complete set $\Psi^{\pm}_{\bm}({\bf 
z},{\bf\zb}) $ of such LLL wave 
functions, we can reconstruct the N-particle coherent states by,
\be{csexp}
\ket{\bz,\pm} &=& C_{\pm}(\bz, \bzb) 
\sum_{\bf m}\ket\bm \bracket{\bm}{\bz,\bzb , \pm}  \\ 
&=& C_{\pm}(\bz, \bzb)  \sum_{\bf m}  \overline\Psi^{\pm}_{\bm}({\bf 
z},{\bf\zb})  \ket\bm
\; , \nonumber 
\ee
and using that the states $\ket\bm$ are normalized, we can express the 
normalization $C_{\pm}(\bz, \bzb)$ as
\be{norm2}
|C_{\pm}(\bz, \bzb)|^{-2} = \sum_{\bf m} |\Psi^{\pm}_{\bm}({\bf 
z},{\bf\zb})|^{2} = \bra{\bz,\bzb \pm} P_{LLL}\ket{\bz ,\bzb ,\pm}
\; .
\ee

The same procedure can now be applied to anyons in the 
LLL provided that we {\em define} the coherent states by the 
projection of the position eigenstates on the LLL. This definition is 
not unique, but can be shown to have several good
properties \cite{Kjonsberg97,Kjonsberg99}.  By substituting \pref{anylll} in the
relations corresponding to
\pref{csexp} and
\pref{norm2} and  redefining the normalization constant by an exponential and
a Jastrow  factor which is common for all the wave functions, we get
\be{acsexp}
\ket{\bz,\nu} = \nN_{\nu}(\bz , \bzb)  \sum_{\bf m}\ket\bm 
S_{\bm}(\bz)  \; ,
\ee
with
\be{annorm}
|\nN_{\nu}(\bz , \bzb)|^{-2} = \sum_{\bf m} |S_{\bm}|^{2} = e^{ \bzb\bz} 
\prod_{i<j}|\za i - \za j|^{-2\nu} \bra{\bz,\bzb, \nu} P_{LLL}
\ket{\bz ,\bzb ,\nu}  \; .
\ee
These expressions are the anyonic counterparts to \pref{symcs} and
\pref{norm}   in the case of bosons and fermions, and the derivation of
the  classical mechanics follows {\em mutatis mutandis}.
The expressions are of course much more complicated than in the boson 
or fermion case, and there is no known analytic expression for the K{\"a}hler potential
except in the case of two particles, where the polynomial part of the 
wavefunction in the relative coordinate $z$ is given by 
\cite{anyonstates}
\be{anwf}
S_{m}(z)= \frac { z^{2m+\nu} } { \sqrt{\pi 
2^{2m}\Gamma(2m+1+\nu) } }\;  .
\ee
The K{\"a}hler potential can then be calculated from \pref{annorm} and 
expressed in terms of a generalized hypergeometric function,
\be{hypgeo}
K(\zb , z) = \hbar\ln\left[\frac 1 {\pi\Gamma(1+\nu)}\ \,   
F_{\hspace{-4mm} 1 \hspace{3mm} 2}\, 
(1; \half +\frac \nu 2 ,1 + \frac \nu 2 ; \frac 
{(\zb z)^{2}} {16})\right] \; . 
\ee
The large  $r$ limit can be obtained from the properties of 
the hypergeometric function, and coincides with the result for bosons 
and fermions. The small $r$ limit can be read off directly from the 
leading term in \pref{annorm},
\be{ansmall}
\lim_{r\rightarrow 0}K(\zb,z) = \hbar\nu\ln\zb z + \frac \hbar
{2(1+\nu)(2+\nu) }
\left(\frac {\zb z}  2 \right)^{2} + const \; ,
\ee 
giving the asymptotic metric 
\be{anymet}
ds^{2} \stackrel {r\rightarrow 0}\rightarrow =
 \frac {2\hbar} {(1 + \nu)(2+\nu)} [\rho^{2} d\theta^{2} + d\rho^{2} ] \; , 
\ee
which interpolates smoothly between the expressions \pref{metlim} for 
bosons and fermions. (Note that the phase space metric found in this way
does not have the same physical dimension as the metric of the space in
which the anyons move. Thus, there is a scale factor $\ell^2/\hbar$ 
between the two metrics, where $\ell$ is the magnetic length.)

In spite of the complicated form of the K{\"a}hler potential $K(\bz, \bzb)$, 
there is a simple property that immediately follows if we write it 
in the following form using the second identity in \pref{annorm},
\be{akah}
K(\bz, \bzb)   =  \hbar[\bzb\bz - 2
\nu\sum_{i<j}\ln |\za i - \za j | + \ln \bra{\bz,\bzb, \nu} P_{LLL}
\ket{\bz ,\bzb ,\nu}]
\; .
\ee
For a translationally invariant state (corresponding to a constant 
magnetic field and no external potential) only the first term can 
depend on the CM coordinate $Z$. Equivalently, we consider the limit 
$z_{i} = Z$ where the positions of all the particles coincide, and get
\be{cmk}
K(Z,\overline Z)   = N\hbar  \overline Z Z \; ,
\ee
corresponding to a single particle of charge $N$, as expected.
Below we shall see that this relation is altered when the particles 
are moving on a sphere, and the corresponding expression will allow us 
to calculate the pertinent phase space volume for particles with 
different statistics.

\vskip 4mm
\noi{\bf B. Identical particles on a sphere}

In this section we will calculate the $N$-particle phase space volume for
particles confined to a finite region. We are interested in the dependence
of the volume on the particle statistics. This extends the previous
discussion of the two-particle case and makes it possible to derive the
(classical) statistical mechanics of the particles. As a convenient
regularization of the system size, we consider a phase space with spherical
geometry\footnote{
The most natural 
choice for a closed two-surface seems to be the torus, since the 
sphere has a finite curvature.  There is, however,  a 
technical difficulty in generalizing the  results in the plane to a 
torus in that the magnetic translation operators are not well defined 
for general translations due to the periodicity 
conditions \cite{Haldane94}. }.

A charged particle moving on a unit sphere,
penetrated by $2j$ units of magnetic flux has a total angular momentum 
$J=j+L$ where $L$ is the orbital angular momentum. The lowest Landau 
level corresponds to $L=0$ and has a $2j+1$ degeneracy \cite{Haldane83}.
Again we can  construct coherent states by acting on an arbitrary minimal 
uncertainty state, that we shall take to be $\ket{j,-j}$, with the 
appropriate group elements of SU(2). We describe the sphere by 
stereographic projection and use a dimensonless complex coordinate, $z$, related to 
the polar angles by, $z =-\tan(\theta/2)e^{-i\phi} $. It is easy to 
show that this $z$ is translated into the dimensionless $z$ introduced 
earlier in \pref{LLL} by the substitution $z\rightarrow \sqrt{\frac 1 {2j}}
z$.  For ease of notation, we shall make this substitution only in the 
final expressions. 

The SU(2) generators in the 
$j$-representation, and the corresponding coherent states are given 
by,

\newcommand{\jpa}[1]{J_{+}^{#1} }
\newcommand{\jma}[1]{J_{-}^{#1} }
\newcommand{\epp}{(1+\zb z)}
\newcommand{\eppa}[2]{(1+\zb_{i} z_{#2})}
\newcommand{\eppj}{(1+\frac {\zb z} {2j})}
\newcommand{\eppja}[2]{(1+\frac {2\zb_{i} z_{#2}} {j})}
\newcommand{\aab}[1]{(1+\overline #1 #1)}

\be{spop}
D(z) = e^{zJ_{+}} e^{\eta J_{0}} e^{-\zb J_{-}} \; ,
\ee
where $J_{i}$ satisfy the standard angular momentum commutation 
relations and $\eta = \ln\epp$. The $D(z)$:s satisfy a multiplication 
rule similar to \pref{dmult}, but here it is sufficient to know the 
overlap,
\be{spoverlap}
\braket z w = [\epp \aab w]^{-j} (1 + \zb w)^{2j} \; .
\ee
Note that in the limit $j = R^{2}/l^{2}\rightarrow\infty$ corresponding to a large 
radius, or a strong magnetic field, we recover,
\be{limit}
\braket z w \rightarrow e^{-\half \zb z -\half \overline w w + 
\zb w}  \; ,
\ee
where we have rescaled $z\rightarrow \sqrt{\frac 1 {2j}}
z$.
We can now immediately take over the results \pref{symcs} and 
\pref{norm} for the  bosonic and fermionic states in the plane ,
\be{spsymcs}
\ket{\bz , \pm} = 
\nN (\bz , \bzb ) \frac 1 
 {\sqrt N!} \sum_{P}\eta_{P}^{\pm}  e^{ \za {i_{P}} \jpa i } 
 \ket{\bf{0}} \; ,
 \ee
and their normalization
\be{spnorm}
|\nN(\bz,\bzb)|^{-2} = \sum_{P,P'}\eta_{P}\eta_{P'}\frac 1 {N!} 
  \bra{{\bf 0}} e^{  \zab{j_{P'}} \jma{j} }  e^{ \za{i_{P}}
\jpa{i}  } \ket{ {\bf 0}} \; .
\ee
Using \pref{spop} and  \pref{spoverlap} we can easily calculate the 
relevant overlap,
\be{spmatel}
\bra{{\bf 0}} e^{\zb{\jma{j} } }  e^{w 
\jpa{i} } \ket{{\bf 0}}  = \delta_{ij} (1 + \zb w)^{2j} \; ,
\ee
so
\be{spnorm2}
|\nN(\bz,\bzb)|^{-2} = \sum_{P}\eta_{P} \prod\limits_i(1 + \zab {i_{P}} \za
i)^{2j}
\; .
\ee

For the case of N coinciding bosons, $\za i = z$,   we immediately get
the following \kah potential
\be{coinbos}
K(z, \zb)  =N\hbar 2j\ln\left(1 + \frac{\zb z}{2j}\right) \; ,
\ee
and the corresponding metric  
\be{spmet}
ds^2 = \frac {2N\hbar} {\eppj^{2}} dz d\zb \; ,
\ee
is just $N$ times that of a sphere. 
Following Manton \cite{Manton93},  
we can use this result to obtain the the volume of the
N-boson phase space\footnote{In \cite{mana} a formula for the volume of the
moduli space for vortices moving on an arbitrary genus Riemannian surface is
derived making extensive use of theorems from complex geometry}.  The essential
observation is that the submanifold spanned by the  configurations of N
coinciding bosons is a complex curve  of degree N in the manifold
$CP_{N}$. The metric \pref{spmet}  immediately gives the volume corresponding to
this complex curve, which can be shown to be N times the area, $A$,
obtained by integrating the fundamental
two form, $\omega$, in \pref{gkh} over a complex line. It then follows from a
general theorem for \kah manifolds that the total volume is given by, 
\be{bphs}
V = \frac 1 {N!} (A)^{N} \; .
\ee
In our case,  using \pref{coinbos} and \pref{gkh}, we get
\be{fundform}
A = \hbar\int_{sph}\, \omega = h 2  j = \frac {\hbar4\pi R^{2}}
{l^{2}} = e\Phi\; ,
\ee
which shows that the area is $h$ times the number of flux quanta 
$\phi_{0}=h/e$ which penetrate the sphere, or equivalently, $\hbar$ 
times the area in units of $l^{2}$.

In the fermion case, we first note from \pref{spnorm2} that the 
normalization can be expressed as a determinant,
\be{fdet}
|\nN(\bz,\bzb)|^{-2} = N! \det\eppa i j^{2j}\; .
\ee
In this case we cannot directly put the particles on top of each 
other (that would give a diverging $\nN$), so we instead put $\za i = 
z + \delta_{i}$ and consider the limit $\delta_{i} \rightarrow 0 $. 
Expanding in $\delta_{i}$, we get,
\be{expdet}
\det\eppa i j^{2j} \simeq \epp^{2jN} \det M_{ij} \; ,
\ee
where
\be{mmat}
 M_{ij} = 1 + \frac {2j} \epp (\zb \delta_{i}+z\overline\delta_{j} ) 
 \; .
\ee
The matrix $M$ is hermitian, so the determinant is real and 
furthermore it has zeros for $\delta_{i} = \delta_{j}$. These 
properties together with power counting is sufficient to establish
\be{mdet}
\det  M_{ij} = C \left(\frac {\zb z} {\epp^{2}} \right)^{\frac{N(N-1)}{2}}
\prod_{i<j}|\delta_{i}-\delta_{j}|^{2} \; ,
\ee
where $C$ is a $z$ and $\delta$ independent constant. 
Combining \pref{expdet} and \pref{mdet} we get, up to a constant,  
the \kah potential
\be{spkah}
K(z,\zb) &= &\ln\epp^{2jN-N(N-1)} + \ln (\zb z)^{\frac{N(N-1)}{2}} \\ 
&\rightarrow &  2jN\left(1-\frac{N-1}  
{2j}\right)\ln\left(1 + \frac{\zb z}{2j}\right) 
- \half {N(N-1)} \ln (\zb z)\; \nonumber  .
\ee
The last term can be removed by a so called \kah gauge transformation, 
and corresponds to redefining the normalization constant by an analytic
factor that  does not contribute to the metric. The first term differs
from the boson case 
\pref{coinbos}  only by the ``reduction'' factor $1-(N-1)/2j$ which equals
one for 
$N=1$, corresponding to a single fermion, and zero for $N=2j+1$ 
corresponding to a filled Landau level. Using the same argument as in 
the boson case, we get the phase space volume,
\be{fphs}
V_{F} = \frac 1 {N!} \left(A - (N-1)h  \right)^{N} \; .
\ee
Again the interpretation is clear - the available phase space for one 
particular fermion is reduced with an amount $h$ by each of the other
particles present in the system. This is consistent with the semi-classical
interpretation, where each quantum state is associated with a phase space
volume $h$. Note that there is a maximum number of particles
allowed, $N=2j+1$, in which case the phase space volume \pref{fphs}
vanishes. This corresponds to the situation where all the lowest angular
momentum states are filled, \ie to a filled lowest Landau level.
Thermodynamically this state is interpreted as being incompressible,
as we will discuss further in Sect.~5. 

Finally  we consider the case of anyons. 
A complete set of LLL anyon wave functions on the sphere corresponding 
to \pref{anylll} in the plane is given by,
\be{spawf}
 \Psi^{\nu}_{\bf m}({\bf z},{\bf\zb})= \prod_{i<j}(\zab i - \zab j)^{\nu}  
\prod_{i}(1+\za i \zab i)^{-j+\frac \nu 2 (N-1)} 
  S_{{\bf m}}({\bf \zb}) \; .
\ee
Following the steps leading from \pref{anylll} to \pref{annorm}, we 
get the anyonic version for \pref{akah},
\be{spkan}
K(\bz , \bzb) &=& \hbar\left(-j + \frac \nu 2(N-1) 
\right)\sum_{i}\ln\left(1 + \frac{\za i \zab i} {2j} \right) \\ &-&
 2\hbar\nu\sum_{i<j}\ln |z_{i} - z_{j}| +  \hbar\ln \bra{\bz,\bzb, \nu} P_{lll}
\ket{\bz ,\bzb ,\nu}
\;  \  . \nonumber
\ee
Note that in this case we do not have any explicit expression for 
the \kah potential corresponding to \pref{spnorm} in the case of 
bosons and fermions. We can, however, again deduce the metric 
corresponding to configurations with coinciding anyons, \ie $z_{i}=z$, from the above 
expression, again noting that  the complicated overlap in
the last term must be independent of $z$, this time because of 
rotational invariance. The  \kah potential becomes,
\be{anykahsp}
K(z,\zb) = \hbar N2j\left(1-\frac{\nu(N-1)}{2j}\right)\ln\left (1+\frac{2\zb 
z}{j}\right) + \ldots 
\ee
which again smoothly interpolates between the bosonic ($\nu =0$) and 
fermionic ($\nu = 1$) results,
\be{afphs}
V_{\nu} = \frac 1 {N!} \left(A - \nu (N-1)h \right)^{N} \; .
\ee

The expressions we have found above for the $N$-particle phase
space volume demonstrates a "classical exclusion principle". Thus, each
new particle introduced in the system will find the available volume
reduced by $\alpha =\nu h$ relative to the previous one. The quantity
$\alpha$, \ie the reduction in phase space volume, can be taken as
defining the classical statistics parameter of the particles. In the
present case it is simply the (dimensionless) quantum statistics parameter
$\nu$ multiplied with Planck's constant $h$. In other cases such a
classical statistics parameter may be possible to define in terms of
reduced phase space volume, even if there is no underlying point particle
description. In the next section we will study such an example.

\vskip 8mm
\noindent{\large\bf 4.  Vortex statistics}
\renewcommand{\theequation}{4.\arabic{equation}}
\setcounter{equation}{0}
\vskip2mm

In the previous sections we have discussed how a phase space description,
with a classical statistics parameter, can be derived from a
(constrained) quantum description. In this section we will consider a
somewhat different system; a classical field theory with soliton
solutions. The system we have in mind is the Chern-Simons Ginzburg-Landau (CSGL) 
theory, originally introduced as a field theory for the quantum Hall
effect \cite{qhalleffth}, with vortices (quasi-particles) as soliton solutions. In
a certain approximation the dynamics can be described in terms of vortex
coordinates alone, and a phase space description can be derived from the
full theory. In this description the vortices will be associated with a
non-trivial classical statistics parameter, and we will show that the value of
this parameter agrees with the value of the (quantum) fractional statistics
parameter usually associated with quasi-particles of the quantum Hall effect.
In this derivation we will make use of the close connection which exists between
the CSGL Lagrangian and the Lagrangian of the relativistic abelian Higgs model
discussed by Samols, Manton and others \cite{Samols92,Manton93,Manton:1997tg}.
The metric of the vortex space is the same in these two cases and it is
K{\"a}hler
\cite{Samols92}. We make use of the results of Manton \cite{Manton93} for the
volume of the
$N$-vortex space, to determine the classical statistics parameter of the
vortices.

The field theory Lagrangian, which describes fields in a 2+1-dimensional
space time, is
\be{oglcs1}
L=\int d^2x[i\hbar\phi^*D_0\phi-\frac{\hbar^2}{2m}|\vec
D\phi|^2-\frac{\lambda}{4}
(|\phi|^2-\rho_0)^2+\mu\hbar\epsilon^{\mu\nu\rho}a_\mu\partial_\nu a_\rho] \; ,
\ee
where $\phi$ is a  complex matter field, $a_\mu$ a Chern-Simons
field,\footnote{
Although the theory is non-relativistic, we choose to use a
relativistic notation for the CS field with $a_0=a^0$, $a_i=-a^i$ (i=1,2) and
$b=\epsilon_{ij}\partial_ia^j$.}
$m$  a mass parameter,
$\lambda$  the interaction strength,  $\rho_0$  the preferred
density of the system and $\mu$ a statistics parameter. 
In this classical field theory $\hbar$ only
plays the role of a dimensional parameter.
For the original Laughlin quantum Hall states
described by the model the statistics parameter takes the values
$\mu=1/[4\pi(2k+1)]$ with 
 integer $k$. $D_0$ and $\vec D$ denote covariant derivatives
\be{coder}
D_0  &=&\frac{\partial}{\partial t}+i a_0 \; , \nonumber \\
\vec D &=& \nabla -i\vec A \; ,
\ee
with
\be{vecpot}
\vec A=\vec a+ \frac{e}{\hbar}\vec A_{ext} \; .
\ee
$\vec A_{ext}$ describes a constant external magnetic field, $B_{ext}$,
which we assume is adjusted to fit the parameter $\rho_0$ so that the
ground state is described by a constant field $\phi$ of density $\rho_0$
with vanishing effective magnetic field, $B=b+\frac{e}{\hbar}B_{ext}=0$. The
physical interpretation is that the system is at  (or close to) the 
center point of a quantum Hall plateau.

It is convenient to change to dimensionless form,
\be{glcs1}
L=\int d^2x[i\phi^*D_0\phi-\frac{1}{2}|\vec
D\phi|^2-\frac{\tilde\lambda}{4}
(|\phi|^2-1)^2+\epsilon^{\mu\nu\rho}a_\mu\partial_\nu a_\rho] \; ,
\ee
where the new Lagrangian is obtained from the original one  by the
substitutions,
\be{rescale}
 &&\phi\rightarrow \sqrt{\rho_0}\,\phi, \,\,\,\,\nonumber \\  
 &&A_0\rightarrow \frac{\hbar\rho_0}{\mu m}A_0,\,\,\,\,\vec A\rightarrow
 \sqrt{\frac{\rho_0}{\mu}}\vec A \; , \nonumber \\
 &&\vec r\rightarrow \sqrt{\frac{\mu}{\rho_0}}\vec r, \,\,\,\, 
t\rightarrow \frac{\mu m}{\hbar\rho_0}t \; , \nonumber \\
 &&L \rightarrow \hbar^2\frac{\rho_0}{m}L \; .
\ee
The single dimensionless parameter in the rescaled Lagrangian is
$\tilde \lambda=\frac{\mu m}{\hbar^2}\lambda $.

There is no independent dynamics associated with the Chern-Simons 
field, since variation with respect to $a_0$ gives a relation between 
magnetic field and (charge) density, which can be written as,
\be{constrain1}
B+(\half \rho-B_{ext})=0 \; ,
\ee
with $\rho=|\phi|^2$. We assume $B_{ext}=1/2$ to give $\rho=1$
and
$B=0$ in the ground state (all fields in dimensionless form). For
finite energy configurations these values are reached asymptotically.
With this assumption the constraint equation \pref{constrain1} for $B$ 
is rewritten as,
\be{Bogomolny1}
B+\half (\rho-1)=0 \; .
\ee 
For stationary states the energy can be expressed as, 
\be{energy}
E=\int d^2x[\frac{1}{2}| 
D_+\phi|^2+\half B+(\tilde \lambda -1)B^2] \; .
\ee
with $D_+\phi=(D_1+iD_2)\phi$. The energy has the lower bound
\be{lowb}
E\geq\int d^2x\half B=N\pi \; ,
\ee
where $N$ is a non-negative integer. For the special value $\tilde
\lambda=1$ of the coupling the lower bound can be saturated. The field $\phi$
then satisfies the linear differential equation
\be{Bogomolny2}
D_+\phi=0 \; .
\ee

The two equations \pref{Bogomolny1} and \pref{Bogomolny2} define (for
$\tilde \lambda=1$) stationary vortex configurations with $N$ vortices of
equal circulation \cite{Taubes80} \footnote{
We have chosen $B_{ext}$ to be positive. With opposite sign the
lower bound is saturated with vortices of opposite circulation.}. 
Since the
configurations (for fixed $N$) are degenerate in energy, the vortices can be
regarded as non-interacting. Gauge-equivalent configurations may naturally
be considered as physically equivalent, and the vortex space can then be
identified with the (moduli) space obtained from the space of field
configurations after the identification of gauge-equivalent 
configurations \cite{Samols92}. A point in vortex space is identified by the
set of
$N$ (unordered) vortex coordinates which corresponds to zeros of the field
$\phi$.

For values of $\tilde \lambda$ close to $1$ the low energy configurations
correspond to slowly moving vortices. A meaningful approximation is then
to impose \pref{Bogomolny1} and \pref{Bogomolny2} as constraints on the field
configurations. The constrained fields describe a system of weakly
interacting vortices. In the $a_0=0$ gauge the Lagrangian takes the 
form,
\be{vortexL}
L=\int d^2x[i\phi^*\dot\phi+\dot{\vec A}\times 
(\vec A-\vec A_{ext})-(\tilde \lambda-1)B^2] - N\pi \; . 
\ee
Due to \pref{Bogomolny1} this Lagrangian is invariant (up to a total time
derivative) under time dependent gauge transformations, $\vec A 
\rightarrow \vec A + \nabla\xi \ ,\  \phi\rightarrow e^{i\xi}\phi$,
and can therefore be interpreted as the vortex Lagrangian defined on the space of
gauge-equivalent field configurations.

It is useful to consider the two fields $\phi$ and $\vec A$ as
components of a complex two-component field \cite{Samols92}, 
\be{u-field}
u=\left( {\matrix{\phi \cr A\cr}} \right) \; ,
\ee
with $A$ as the complex field $A=A_1+iA_2$. A hermitian scalar product
is introduced as
\be{scalprod}
\left\langle {u} \mathrel{\left | {\vphantom {u v}} \right.
\kern-\nulldelimiterspace} {v} \right\rangle =\int d^2x\, u^{\dagger}v
\; . 
\ee
With this notation, up to total time derivatives the kinetic term in the
Lagrangian (\ie the part with time derivatives) can be written as
\be{kinetic}
T=i\langle u-u_{ext}| \dot u \rangle \; ,
\ee
with $u_{ext}=(0,A_{ext})$.

The vortex configurations are described by a set of vortex coordinates
$x=\{x_1,x_2,...,x_n\}$, with $x_i$ corresponding to two real or one
complex coordinate of vortex $i$. The precise form of the multi-vortex
configurations for given coordinates is not known, but their existence
is \cite{Taubes80}. The kinetic term for these
configurations can be written
\be{kinetic2}
T={\cal A}_i(x) \dot x_i \; ,
\ee
where the new vector potential is
\be{vecpot2}
{\cal A}_i &=&i\langle u-u_{ext}|\partial_i u\rangle \nonumber \\
&=& i\langle u|\partial_i u\rangle-\partial_i\langle u_{ext}| u\rangle \; ,
\ee
and where the differentiation now is with respect to the vortex
coordinates. The corresponding field tensor which defines the symplectic
form and the phase space structure of the vortex space is
\be{fieldtensor}
{\cal F}_{ij}=\partial_i {\cal A}_j-\partial_j {\cal A}_i = -2\Im\langle\partial_i
u|\partial_j u \rangle \; .
\ee

The vortex space is the space of gauge equivalent field
configurations which satisfy \pref{Bogomolny1} and \pref{Bogomolny2}. Let us
re-consider this gauge equivalence in terms of the complex fields $u$. The
infinitesimal gauge transformations have the form
\be{gaugetr}
\delta_vu=(i\chi\phi,2\partial_{\bar z}\chi) \; ,
\ee
with $\chi$  a real function, so the vectors $\delta_v  u$ define a real
vector space. This  can be extended to a complex vector
space if $\chi$ is allowed to be complex. We will refer to the corresponding
transformations as complex gauge transformations, with $\Gamma(u)$ as the
projection onto this complex subspace, and we will refer to the directions
defined by the gauge transformations
\pref{gaugetr}  as vertical directions. The orthogonal directions (horizontal
directions) are defined by variations in the field
$u$,
\be{delta-hu}
\delta_h u=(\delta\phi,\delta A) \; ,
\ee
which satisfy $\langle\delta_v
u|\delta_h u \rangle=0$, implying
\be{horizontal}
2\partial_{z}\delta A+i\phi^*\delta\phi=0 \; .
\ee
The real part is
\be{real}
\nabla\cdot\delta \vec A+\frac{i}{2}(\phi^*\delta\phi-\phi\delta\phi^*)=0 \; ,
\ee
and the imaginary part is
\be{imaginary}
\delta B+\half\delta \rho=0 \; .
\ee
Changes in the fields which follows from variations in the vortex coordinates
will automatically satisfy the second equation, \pref{imaginary}, due to the
constraint \pref{Bogomolny1}. The first equation, \pref{real}, {\em can} be
satisfied provided we make use of the freedom to include {\em real} gauge
transformations in the variations of the fields. Thus, we may assume both these
equations, or equivalently \pref{horizontal} to be satisfied when
$\delta A$ and $\delta \phi$ are replaced by the corresponding derivatives with
respect to vortex positions.

We introduce $\Pi(u)=I-\Gamma(u)$ as the projection onto the horizontal
directions and $D_i=\Pi\partial_i$ as the projected derivative. With the
assumption that the gauge condition \pref{real} is satisfied the vector
potential ${\cal A}_i$ can be written as
\be{vecpot3}
{\cal A}_i=i\langle u-u_{ext}|D_i u\rangle \; . 
\ee
It transforms under a (complex) gauge transformation $\chi$  as 
\be{transf}
{\cal A}_i\rightarrow {\cal A'}_i={\cal A}_i-\partial_i\Theta \; . 
\ee
with 
\be{theta}
\Theta=\int d^2x[\phi^*\phi+2i\partial_{\bar z}(A^*-A^*_{ext})]\chi 
\; .
\ee
This means that ${\cal F}_{ij}$ is invariant under complex gauge transformations
and can be written as
\be{fieldtensor2}
{\cal F}_{ij}= -2\Im \langle D_i u| D_j u \rangle \; , 
\ee
since the difference between $\partial_i u$ and $D_i u$ can (locally) be
eliminated by a (vortex-position dependent) gauge transformation.

We may consider the quantity
\be{metric2}
\eta_{ij} =\langle D_i u| D_j u \rangle
\ee
as defining a hermitian metric tensor for the vortex space. It is obtained from
the scalar product \pref{scalprod} by projection on the horizontal directions.
The  tensor ${\cal F}_{ij}$, which is derived from the kinetic
part of the Lagrangian, and which defines the phase space structure of the vortex
space, can now be identified with the imaginary part of this metric
tensor. 

The metric \pref{metric2} is relevant also for the relativistic abelian Higgs
model, as we shall now demonstrate. 
The Lagrangian of this model has the
form
\be{Higgs}
L=\int d^2x
[\half(D_{\mu}\phi)^* D^{\mu}\phi
-\frac{1}{4}F_{\mu\nu}F^{\mu\nu}-\frac{2\tilde \lambda-1}{8}\;
(|\phi|^2-1)^2] \; .
\ee
It is quadratic in time derivatives and has the Chern-Simons field replaced
by a Maxwell field. The energy has the same lower bound \pref{lowb} as the
GLCS model, and for $\tilde \lambda=1$ this lower limit  is saturated if both
the equations \pref{Bogomolny1} and \pref{Bogomolny2}, known as the Bogomolny
equations, are satisfied. Thus,  if these equations are used to define the
$N$-vortex space, the vortex space is the same for the two models. However,
the kinematics, as defined by the kinetic part of the Lagrangian is not the
same in the two cases. The non-relativistic model is linear in time
derivatives, which means that the vortex space has the character of a phase
space, while the relativistic model is quadratic in time derivatives, which
gives the vortex space the character of a configuration space.

Expressed in the $A_0=0$ gauge and constrained by the Bogomolny equations,
the Lagrangian of the Higgs model has the form
\be{HiggsLagrange}
L'=T'-V \; ,
\ee
with
\be{TV}
T'&=&\half \int d^2x [\dot \phi^* \dot \phi + \dot {\vec A}\cdot  \dot {\vec
A}]\; , 
\nonumber
 \\ V&=&\int d^2x (\tilde \lambda-1)B^2 + N\pi  \; ,
\ee
and the fields constrained by Gauss' law
\be{Gauss}
\nabla\cdot {\dot {\vec A}}+\frac{i}{2}(\phi^*\dot\phi-\dot\phi^*\phi)=0 \; .
\ee  The potential $V$ is the same, but the kinetic term $T'$ is different from
that of the CSGL model. We note that the constraint
\pref{Gauss} corresponds to the real part
\pref{real} of the condition for motion in horizontal direction. Also the
imaginary part
\pref{imaginary} is satisfied due to the Bogomolny equations. The constraint on
the motion, given by Gauss' law, can be expressed in terms of the projection
$\Pi$ and the field
$u$ as
\be{T'} T'&=&\half \langle\dot u|\Pi|\dot u\rangle \nonumber \\ &=&\half\dot
x_i\dot x_j\langle D_i u|D_ju\rangle \; .
\ee

The kinetic term $T'$ is different, but related to the kinetic term $T$ of the
CSGL model. $T'$ is determined by the real part, whereas $T$ is determined (up
to a gauge transformation) by the imaginary part of the same hermitian metric
\pref{metric2}. This metric has been examined in some 
detail by Samols \cite{Samols92} for the case of the abelian Higgs model. It is
a K{\"a}hler metric, which is conveniently expressed in terms of complex
vortex coordinates as
\be{metric3}
ds^2=-2i{\cal F}_{\zb_iz_j} d\zb_i dz_j \; ,
\ee
with
\be{Fij}
{\cal F}_{\zb_iz_j}=i\left[\langle {{D_{\zb_i}u}|{D_{z_j}u}}\rangle -
\langle {{D_{z_j}u}|{D_{\zb_i}u}}\rangle \right] \; . 
\ee
The corresponding K{\"a}hler 2-form is
\be{2-form}
\omega = -{\cal F}_{\zb_iz_j}d\zb_i \wedge dz_j \; .
\ee

The K{\"a}hler form determines the symplectic structure and the volume of the
vortex space. This (2$N$-dimensional) volume is the same whether the vortex space
is considered as a configuration space (\ie with the volume determined by the
real part of the metric) or as a phase space (with the volume determined by the
imaginary part), and has been calculated by Manton \cite{Manton93} for
$N$ vortices on a sphere. The result is \footnote{
Manton's result has been
changed with a factor $(2\pi)^N$ to fit the definition of the metric in this
paper. We also express the $N$-vortex volume in terms of the single-vortex
volume $A$ rather than the area of the sphere. This gives a factor $N-1$
instead of $N$ in the second term of Eq.\pref{volume}.}  
\be{volume}
V_N={1\over{N!}}(A-8\pi^2 (N-1))^N \; ,
\ee
with $A$ as the volume (area) of the one-vortex space.
The volume
\pref{volume} has the same form as discussed in Sect. 3 for $N$ identical
particles with non-trivial classical statistics. With the vortex space
interpreted as a phase space, the statistics of the vortices can be extracted
from the reduction term
$8\pi^2 N$. In order to do so correctly we have to re-write the volume
\pref{volume} in dimensional form. The phase space volume is determined by the
Lagrangian
\pref{glcs1}, and as follows from the transformations \pref{rescale} the phase
space dimensions are correctly re-introduced by the substitutions
$A\rightarrow\mu \hbar A$ and $V_N\rightarrow(\mu\hbar)^NV_N$. In dimensional
form the expression for the $N$-vortex volume is
\be{volume2} 
V_N={1\over{N!}}(A-4\pi\mu h (N-1))^N \; ,
\ee
and the classical statistics parameter as determined by the reduction in
available phase space due to the presence of other vortices therefore is
\be{statfac}
\alpha=4\pi\mu h\equiv g h \; .
\ee
We can interpret $g$, the classical parameter divided by $h$, as the
dimensionless quantum statistics parameter.
The value
$g= 4\pi\mu$ agrees with the value of the statistics parameter as determined
from  Berry phase calculations with Laughlin wave functions \cite{Arovas84}, 
or from the properties of vortices in the CSGL model \cite{qhalleffth}.

\vskip 8mm
\noindent{\large\bf 5.   Statistical mechanics  }

\newcommand{\bea}{\begin{eqnarray}} 
\newcommand{\eea}{\end{eqnarray}}
\newcommand{\e}{\varepsilon} 
\newcommand{\D}{\partial}

\renewcommand{\theequation}{5.\arabic{equation}}
\setcounter{equation}{0}
\vskip3mm
\noi{\bf A. Entropy and pressure of identical particles}

We have already emphasized that even if the classical equations 
of motion, and thus the {\em classical dynamics}, does not depend on the 
classical statistics parameter $\alpha$, the {\em statistical 
mechanics} (and thus the thermodynamics), does. In this section we 
demonstrate this by first calculating the entropy and pressure in the 
two model systems considered in sections 3 and 4, charged particles 
in the lowest Landau level, and vortices in the  CSGL model 
respectively.  The results of this calculation fit nicely into the 
general framework of ``fractional exclusion statistics'' for 
particles with degenerate energy levels, and we briefly review the 
basics of this topic before  further discussing  our results. 

We assume the interaction strength to be $\tilde \lambda =1$ for the CSGL
theory. The vortex system is then degenerate in energy. That is also the
case for a system of anyons in the lowest Landau level. Thus, both these
systems have the special property that the energy does not depend on
the state, but only on the number of particles. This means that the
statistical mechanics is determined by the phase space volume
$V_N$, which has been determined in the previous sections, and by the energy
$E_N$. The classical partition function is simply the total number  of
states,
$ { V_{N}}/ h^{N}$ multiplied  with the Boltzmann factor, \ie 
\be{part}
Z_{N} =\frac { V_{N}} {h^{N}  } e^{-\beta E_{N}} \; ,
\ee
The following simple expressions for the free energy 
$F$ and the entropy $S$, immediately 
follow,
\be{freee}
F &=& E_{N}-T\ln (V_{N}/h^{N}) \; , \\
S &=& \ln (V_{N}/h^{N})\;  . \nonumber 
\ee
where  is Boltzmann's constant is set to unity. The pressure is usually
defined by
$P = -(\partial F/\partial {\cal  V})_{T}$, where ${\cal V}$ is the
volume of real space, but in  the systems we have considered the real
two-dimensional space where  the particles or vortices move is proportional to
the phase space, so we simply define the pressure as,
\be{pres}
P = -\left(\frac {\partial F} {\partial A} \right)_T = 
  T\;\frac { \partial \ln {(V_N/h^N) }  }  {\partial A} 
 \; ,
\ee
where $A=V_{1}$ is the phase space volume for  a single particle. 
Substituting the results \pref{afphs} or \pref{volume2}, we get
\be{ent}
S &=& N\ln(1 - \alpha\rho) + N\ln \frac A h - N\ln N + N \; , \\
\beta P &=& \frac \rho {1 - \alpha \rho} \; , \label{pres2}
\ee
where  $\alpha=\nu h$ or $gh$, and where we have introduced the
classical phase space density $\rho =N/A$  and neglected the difference between
$N$ and
$N-1$, which is  irrelevant in the thermodynamic limit. 

The expression \pref{pres2} shows that there is a maximum density
$\rho =1/\alpha$ allowed by the system, which corresponds to an
infinite pressure and therefore to an incompressible state.
For the phase space volume this means $V_{N} = 0$, \ie there is no
available phase space volume for any new particle added to the system. For
the anyon system this situation corresponds to a completely filled Landau
level. What is unusual about this is that the blocking, which can be
interpreted as representing a generalized Pauli principle, shows up not
only in the quantum  but also in the classical description of
the system. 

\vskip 4mm
\noi
{\bf B. Exclusion statistics and the classical limit} 

The generalization of the Pauli exclusion principle introduced by Haldane
\cite{Haldane91}, usually called {\em exclusion statistics}, states
that in the presence of particles in a set of given quantum states, the
number of available one-particle states for any new particle added to the
system is reduced. More
precisely,  the addition of
$\Delta N$ particles changes the number of available states, $d_N$ 
according to 
\be{exstat}
\Delta d_N = - g\Delta N \; ,
\ee
where $g$ is the exclusion statistics parameter. The statistical weight,
or number of states available for the full $N$-particle system, is given by
the formula
\be{W}
W_N=\frac{(G+(1-g)(N-1))!}{N!(G-gN-(1-g))!} \; ,
\ee 
where $G$ is the number of single-particle states. Clearly \pref{exstat}
and
\pref{W} reduces to standard expressions when
$g=0$ (for bosons, with no exclusion) and
$g=1$ (for fermions, with total exclusion). There exist some
(theoretical) realizations of exclusion statistics for particles in one
dimension (\ie with two-dimensional phase space) for $g$ different from
these two values. One particular case is the system of anyons confined to
the lowest Landau level, which we have already considered
\cite{Haldane91}. In that case the exclusion statistics parameter $g$ is
identical to the anyon statistics parameter $\nu$. 

The statistical mechanics of particles with exclusion statistics can be
derived from the statistical weight \pref{W} when the total energy can be
written as a sum of single-particle energies and \pref{W} is 
applied separately to (single-particle) energy levels
\cite{Isakov94,dVO94,Wu94}. The result for the entropy is
\be{exent}
S &=& \sum_{k} D_{k}  \{ [1+(1-g)n_{k}] \ln [1+(1-g)n_{k}] \\ &+& 
(1-gn_k) \ln (1-gn_{k}) - n_{k} \ln n_{k} \} \; , \nonumber
\ee 
where the sum runs over single-particle energy states. $D_{k}$ is the
degeneracy of  the $k$-th level and  the {\em quantum distribution
function}, $n_{k}$,  is the average occupation number of the state $k$. 

Since each quantum state occupies the phase space volume $h^{\cal D}$,
with $2{\cal D}$  the dimension of the single-particle phase space,  we
can relate $n$ and
$\rho$ in  the semiclassical limit by $n=\rho h^{\cal D}$. In the Boltzmann
limit,  $h \rightarrow 0$ and $n\rightarrow 0$, all 
dependence on $g$ in \pref{exent} goes away. If we, however, 
define the classical physics by the double limit  $h \rightarrow 
0$, $g \rightarrow \infty$ and $gh^{\cal D} \rightarrow \alpha$, where 
$\alpha$ is interpreted as a classical statistics parameter\footnote{
Such a
way of taking the classical limit is well-known from other contexts. Thus,
a charged particle can in the quantum mechanical description be characterized
by a dimensionless charge $g=q/\sqrt{\hbar c}$, where $q$ is the physical
charge. (For $q=e$ we have $g^{2} = 4\pi\alpha$, with $\alpha$ the 
fine structure constant.)
With $g$ fixed the charge $q$ depends on $\hbar$ and vanishes in
the limit $\hbar\rightarrow0$. However, if the classical limit is taken as
$\hbar\rightarrow0,g\rightarrow\infty$ with $g\sqrt{\hbar c}\rightarrow
q$ the dimensional charge $q$ survives the classical limit. }
\pref{exent} gets a nontivial  limit of
\be{cexent}
S = \sum_{k} D_{k}h^{\cal D}\left[ \rho_k\ln(1-\alpha\rho_k) -
\rho_k\ln(\rho_k h) +\rho_k 
\right]
\; .
\ee
If we further specialize to the case of 
fully degenerate states in a two-dimensional phase space,  
where the sum is simply replaced by the total number of 
available single-particle states, $G = A/h$, and where $\rho_k$
is replaced by $N/A$, we exactly regain \pref{ent}. This demonstrates that
the classical statistical mechanics discussed in the previous section can
be regarded as a special limit of exclusion statistics, different from the
Boltzmann limit.  

An alternative way to see the correspondence is  to start from the the
equation of state for exclusion statistics particles with  the same
energy,
\be{exeos}
\beta P = \frac{G}{\cal V} \ln \left(  1 + \frac{n}{1- g n} 
\right) \; ,
\ee 
where $n=N/G$. Introducing the density $\tilde\rho = N/{\cal V}$ and
taking the  double limit defined above we get,
\be{exeos2}
\beta P = \frac{\tilde\rho}{1-\alpha \tilde\rho \frac{\cal V}{V_{1}}} \; .
\ee
If we identify the the physical volume ${\cal V}$ with the one 
particle phase space volume $A$, so that $\tilde\rho = \rho$, we 
reproduce \pref{pres2}. 

In the Appendix it is shown that even in a non-degenerate case, with
particles in a harmonic oscillator potential, the classical statistical
mechanics, defined as in Sect. 5A, coincides with that of exclusion
statistics when the classical limit is taken in the way discussed above.
Clearly what is important for the connection with exclusion
statistics is the two defining relations \pref{exstat} and \pref{W}  which
determine the number of states in the system. In the classical description
they are represented by the expressions for the phase space volume, and by
taking the limit $h\rightarrow0,\ g h^{\cal D}\rightarrow\alpha$ it is
straightforward to demonstrate that \pref{exstat} and \pref{W} reproduce
the expressions for the phase space volume derived in Sects. 3 and 4.  

\vskip 8mm
\noindent{\large\bf 6. Discussion  }

In this paper we have described a way to encode the particle statistics in
the classical Lagrangian of a many particle system. The important point is
that the Lagrangian includes more information about the system than just
the classical equation of motion. It also gives information about the
volume of the phase space, which in the quantum description
corresponds to the number of states. If the $N$-particle volume can be
determined as a function of the single particle volume, a classical
statistics parameter can be defined as the reduction in available phase space
volume for one particle  by the presence of the others. Viewed in this
way this classical statistics can be regarded as an analogue of
exclusion statistics. In the specific examples we have considered, this
relation can be made more specific and the classical statistical mechanics
derived from this can be seen as a special way to take classical limit of
exclusion statistics.

To make this idea more precise we have considered cases where the
classical mechanics can be derived from the quantum description by
constraining the motion in Hilbert space to (generalized) coherent states.
For bosons, fermions and even anyons with a two-dimensional phase space
the Lagrangian can be derived and the phase space volume can be
calculated. The dimensional, classical statistics parameter, defined as
the volume occupied by each particle present, in these cases are simply
the dimensionless quantum statistics parameter multiplied with Planck's
constant $h$. In another example, vortices in CSGL theory, there is no
such underlying point particle description, but a similar classical
Lagrangian can be found and the classical statistics parameter can be
related to the coupling of the Chern-Simons term.

There are several interesting questions raised by this description:\\
$\bullet$ Is the "classical fermion" description useful in some cases? This
description would correspond to retaining the fermions' ability to occupy
phase space, but otherwise treat them as classical particles. (A classical
electron would then be characterized both by a charge and a (classical)
statistics parameter.) Can the description give a useful approximation
for other objects, like vortices in superfluids or superconductors?\\ 
$\bullet$  In the examples we have studied the phase space is two-dimensional,
but the formalism (like for exclusion statistics) does not seem to depend
in any crucial way on dimension. 
Are there non-trivial  higher-dimensional examples?
(Fermions in two and three dimensions can certainly be represented like
this.)\\
$\bullet$  What about quantizing such a classical theory? In the cases we
have studied, with a K{\"a}hler metric defined on phase space, a
quantum description can presumably be derived in a unique way by use
of analyticity properties. When regarded as a "re-quantization" of
the system, how does it relate to the original quantum description.
What would in particular the quantum description of the CSGL
vortices be?

All these questions seem to merit further
investigation.       

\vskip 4mm
\noindent{\large\bf Appendix: Harmonic oscillator potential}
\renewcommand{\theequation}{A.\arabic{equation}}
\setcounter{equation}{0}
\vskip2mm

In this Appendix we consider the statistical mechanics of particles in a
harmonic oscillator potential. The particles are "classical anyons" in
the sense discussed in Sect. 3, \ie the one derived from  quantum
 anyons in the lowest Landau level. The system can also be
interpreted as a coherent state representation of particles in a
one-dimension harmonic oscillator potential, in a form interpolating
between bosons and fermions. We calculate the partition function of the
$N$-particle system and show that this is related to the partition
function of a (quantum) system of particles with exclusion statistics in a
harmonic oscillator potential by the same correspondence as obtained in
Sect. 4.

The wave functions of the lowest Landau level have the form
\be{psiLLL}
\psi(\bz,\bzb)=\prod\limits_{i<j}(\zb_i-\zb_j)^\nu
f(\bzb)\;e^{-\half \bzb\bz} \; ,
\ee
with $f(\bzb)\equiv f(z_1,...,z_N)$ as a general anti-analytic function of the complex
particle coordinates. It is assumed to be symmetric in the variables. We introduce
analytic basis vectors by
\be{anbas}
\left\langle {\bz} \mathrel{\left | {\vphantom {{z\_
1,...z\_ N} \psi }} \right. \kern-\nulldelimiterspace} {\psi }
\right\rangle =f(\bzb)  \; .
\ee
The basis vectors $|z_1,...,z_N\rangle$ are not normalized, but we assume
$\nu$ to be chosen such that they are regular and non-vanishing at
points of coincidence of particle positions. Normalized vectors are
introduced by
\be{normA}
|\psi_{\bz,\bzb}\rangle&=&\nN_{\bz,\bzb}\ket\bz \; , \nonumber
\\ |\nN_{\bz,\bzb}|^{-2}&=&\langle \bz|\bz\rangle \; .
\ee
Defined in this way $|\nN_{\bz,\bzb}|^{-2}$ is a regular function with
no zeros anywhere in $N$-particle space, and the K{\"a}hler potential
$K=\ln|\nN|^{-2}$ is a regular function everywhere.

The Hamiltonian depends on two frequencies, the cyclotron frequency
$\omega_c$ determined by the external magnetic field and the frequency
$\omega_0$ of the additional harmonic oscillator potential. When acting
on the anti-analytic part $f(\bzb)$ of the wave functions of LLL, the
Hamiltonian has the form
\be{Ham} 
H &=&\hbar (\omega_t-\omega_c)\sum\limits_i\zb_i\partial_{\zb_i}+\hbar
\omega_t\left[\frac{\nu}{2}N(N-1) +\frac{N}{2}\right] \\
&=&  \hbar \omega\sum\limits_i\zb_i\partial_{\zb_i} + V_N^0 \; ,
\ee 
with $\omega_t=\sqrt{\omega_c^2+\omega_0^2}$,
$\omega=\omega_t-\omega_c$ and  $V_N^0$  
the quantum mechanical ground state energy 
\be{VN0}
V_N^0=\hbar \omega_t\left[\frac{\nu}{2}N(N-1)
+\frac{N}{2}\right] \; .
\ee
For a system of particles in a
one-dimensional harmonic oscillator potential the Hamiltonian is essentially
the same, except that it depends on a single frequency $\omega_0$,
\be{Ham2} H=\hbar
\omega_0\sum\limits_i\zb_i\partial_{\zb_i}+\hbar\omega_0\left[\frac{\nu}{2}N(N-1)
+\frac{N}{2}\right] \; .
\ee 
Thus the difference between these two cases is only an overall
$N$-dependent shift of the energy spectrum.

The energy  of the classical description is determined by the matrix
elements of the Hamiltonian \pref{Ham},
\be{apppot} V(\bz)&=&\langle
\bz|H|\bz\rangle|\nN_{\bz,\bzb}|^2 \nonumber \\
&=&\left[\left\{\hbar\omega\sum\limits_i\zb_i\partial_{\zb_i} +\hbar
\omega_t\left[\frac{\nu}{2}N(N-1)
+\frac{N}{2}\right]\right\}|\nN_{\bz,\bzb}|^{-2}\right]|\nN_{\bz,\bzb}|^2
\nonumber\\ 
&=&\hbar\omega\sum\limits_i\zb_i\partial_{\zb_i} \ln
|\nN_{\bz,\bzb}|^{-2} +\hbar\omega_t\left[\frac{\nu}{2}N(N-1)
+\frac{N}{2}\right] \; , \nonumber \\
\ee
and the $N$-particle partition function is
\be{partition}
Z_N=\frac{1}{h^N}\int \frac{\omega^N}{N!} e^{-\beta V} \; ,
\ee
where $\omega$ is the symplectic form
\be{omega}
\omega=- f_{\bar z_i z_j}d\bar z_i\wedge dz_j \; ,
\ee
with
\be{fij}
f_{\bar z_i z_j}=i\hbar\partial_{\bar z_i}\partial_{z_j}
\ln|\nN_{\bz,\bzb}|^{-2} \; .
\ee
The form of the energy  makes it possible to evaluate the integrals in the
expression for the partition function. We write it as
\be{partition2}
Z_N=\frac{e^{-\beta V_N^0}}{\pi^N N!}\; \epsilon_{ij...k}\int
d^2z_1...d^2z_N
\;[\partial_{\bar z_i}\partial_{z_1}
\ln|\nN_{\bz,\bzb}|^{-2}...\partial_{\bar z_k}\partial_{z_N}
\ln|\nN_{\bz,\bzb}|^{-2}]
\nonumber \\
\times\exp\{-\beta[\hbar\omega\sum\limits_iz_i\partial_{z_i}
\ln |\nN_{\bz,\bzb}|^{-2} ]\} \; .
\nonumber \\
\ee
The partition function can be rewritten as
\be{partition3}
Z_N=\frac{1}{(-\beta\hbar\omega)}\frac{e^{-\beta
V_N^0}}{\pi^N N!}\; \epsilon_{ij...k}\int d^2z_1...d^2z_N
\;\frac{1}{z_1}\partial_{\bar z_i}\large[
\ln|\nN_{\bz,\bzb}|^{-2}...  \nonumber \\ 
 \times\partial_{\bar z_k}\partial_{z_N}
\ln|\nN_{\bz,\bzb}|^{-2}\exp\{-\beta(\hbar\omega\sum\limits_iz_i\partial_{z_i}
\ln |\nN_{\bz,\bzb}|^{-2})\}\large] \; ,
\nonumber \\
\ee
and by use of the identity
\be{ident}
\frac{1}{z_1}\partial_{\bar z_i}=\partial_{\bar z_i}\frac{1}{z_1}-\pi 
\delta(z_1)\delta_{i1}
\ee
the integration over $z_1$ can be performed
\be{partition4}
Z_N=\frac{\pi}{\beta \hbar\omega}\frac{e^{-\beta
V_N^0}}{\pi^N N!}
\;\epsilon_{j...k}\int d^2z_2...d^2z_N
\;[\partial_{\bar z_j}\partial_{z_2}
\ln|\nN_{\bz,\bzb}|^{-2}...
\nonumber \\
\times\partial_{\bar z_k}\partial_{z_N}
\ln|\nN_{\bz,\bzb}|^{-2}]\exp\{-\beta[\omega\sum\limits_iz_i\partial_{z_i}
\ln |\nN_{\bz,\bzb}|^{-2} ]\} \; .
\nonumber \\
\ee
The $(N-1)$-particle integral in this expression is of the same form as the original
$N$-particle integral, and by repeating the procedure $N$ times we get the
following simple expression for the partition function
\be{partition5}
Z_N&=& \frac{1}{(\beta \hbar\omega)^N N!} e^{-\beta
V_N^0}
\nonumber \\ &=& \frac{1}{(\beta \hbar\omega)^N N!}
\exp\{-\beta \hbar
\omega_t\left[\frac{\nu}{2}N(N-1) +\frac{N}{2}\right]\} \; .
\ee

The classical expression for the partition function can be compared with the partition
function of the quantum system
\be{qpart}
Z_N=Tr\;e^{-\beta H} \; ,
\ee
with $H$ given by \pref{Ham}. This expression is easily evaluated, since it can be
written as
\be{qpart2}
Z_N&=&e^{-\beta V_N^0}\sum\limits_{l_1=0}^\infty\sum\limits_{l_2=l_1}^\infty...
\sum\limits_{l_N=l_{N-1}}^\infty\;e^{-\beta
\hbar\omega\sum\limits_i l_i}
\nonumber \\ &=&e^{-\beta V_N^0}\left[\prod\limits_{n=1}^N(1-e^{-n\hbar\beta
\omega}\right]^{-1} \; .
\ee
This expression shows that in the limit $\hbar\rightarrow0$, with $\hbar \nu$ fixed, the
partition function \pref{qpart2} of the quantum system coincides with the classical
partition function \pref{partition4}. (Note however that  the classical function
depends on $\hbar$ explicitly, not only through the statistics factor
$\alpha=h\nu$, due to the contribution from the ground state energy.) 

It is well known that the system of particles in the lowest Landau level
can be regarded as a special realization of exclusion statistics
\cite{Haldane91}, and the correspondence between the two partition
functions discussed here is therefore essentially the same as the
correspondence between the classical statistical mechanics and the
statistical mechanics of particles with exclusion statistics discussed in
Sect. 4. If we use the harmonic oscillator as a volume regulator the
relation between the discussion in this Appendix and in Sect. 4 becomes
even more direct. The thermodynamic limit is here taken by interpreting
the limit
$\omega_0\rightarrow 0$ in a specific way \cite{regularization}. For the
quantum case the harmonic oscillator regulator has been used in
\cite{dVO94}, and the expressions for the entropy and equation of state of
anyons in the LLL were found in this way. Due to the correspondence
between the quantum and classical descriptions, the thermodynamic limit
of the classical functions with the harmonic oscillator regularization,
will be identical to the corresponding functions of Sect. 4. That is what
should be expected, since for the thermodynamic limit it should be of no
significance whether volume regularization is done by confinement to a
sphere or by confinement in a harmonic oscillator potential.

\vskip 4mm
\noindent{\large\bf Acknowledgment}

Serguei Isakov acknowledges the support received through a NATO Science
Fellowship granted by the Norwegian Research Council. S.I. also
appreciates warm hospitality of NORDITA during his stay there in the
summer of 1999, where part of this work was done.

Three of us (T.H.H., S.I. and J.M.L.) would also like to thank the
Department of physics, Norwegian University of Science and Technology
(Trondheim) for the inspiring atmosphere of the Workshop on
low-dimensional physics in June 1999.

\eject

\end{document}